\documentclass[conference]{IEEEtran}
\IEEEoverridecommandlockouts
\usepackage{cite}
\usepackage{amsmath,amssymb,amsfonts}
\usepackage{algorithmic}
\usepackage{graphicx}
\usepackage{textcomp}
\usepackage{xcolor}
\usepackage{float}
\usepackage{booktabs}
\usepackage{url}
\def\BibTeX{{\rm B\kern-.05em{\sc i\kern-.025em b}\kern-.08em
    T\kern-.1667em\lower.7ex\hbox{E}\kern-.125emX}}
\newcommand{\eps}{\varepsilon}
\newcommand{\OPT}{\mathsf{OPT}}
\newcommand{\OPTW}{\mathsf{OPT_W}}

\begin{document}

\title{Mass Exit Attacks on the Lightning Network \thanks{A publication derived from this work was presented at IEEE ICBC 2023.}}

\author{\IEEEauthorblockN{Cosimo Sguanci}
\IEEEauthorblockA{
\textit{University of Illinois at Chicago}\\
Chicago, USA \\
csguan2@uic.edu}
\and
\IEEEauthorblockN{Anastasios Sidiropoulos}
\IEEEauthorblockA{
\textit{University of Illinois at Chicago}\\
Chicago, USA \\
sidiropo@uic.edu}}

\maketitle

\begin{abstract}
The Lightning Network (LN) has enjoyed rapid growth over recent years, and has become the most popular scaling solution for the Bitcoin blockchain.
The security of the LN relies on the ability of the nodes to close a channel by settling their balances,
which requires confirming a transaction on the Bitcoin blockchain within a pre-agreed time period.

We study the susceptibility of the LN to mass exit attacks in case of high transaction congestion, in the presence of a small coalition of adversarial nodes that forces a large set of honest users to interact with the blockchain.
We focus on two types of attacks: (i) The first is a zombie attack, where a set of k nodes become unresponsive with the goal of locking the funds of many channels for a period of time longer than what the LN protocol dictates.
(ii) The second is a mass double-spend attack, where a set of k nodes attempt to steal funds by submitting many closing transactions that settle channels using expired protocol states;
this causes many honest nodes to have to quickly respond by submitting invalidating transactions.

We show via simulations that, under historically plausible congestion conditions, 
with mild statistical assumptions on channel balances,
both attacks can be performed by a very small coalition. To perform our simulations, we formulate the problem of finding a worst-case coalition of k adversarial nodes as a graph cut problem.
Our experimental findings are supported by theoretical justifications based on the scale-free topology of the LN.
\end{abstract}


\section{Introduction}

The Lightning Network (LN) \cite{poon2016bitcoin} is emerging as the most popular layer-2 scaling technology for Bitcoin.
We model the LN by a graph $G$;
every vertex can be assumed to be a user
and every edge is a bi-directional payment channel between two nodes. The LN is implemented via a cryptographic protocol executed by the nodes. Effectively, pairs of nodes implement a payment channel that allows them to exchange promises to pay a certain amount to each other. These promises are valid, as long as they are redeemable on layer-1, i.e. the Bitcoin blockchain.
The LN is \textit{permissionless}, that is, any user can join the protocol pseudonymously.
Therefore, users can have arbitrarily adversarial behavior, so the LN must be able to handle byzantine failures.
This implies that each node in a channel $e\in E(G)$ must be able to unilaterally close $e$ by performing layer-1 transactions.
However, in order to prevent double-spend, the protocol must prohibit a node from closing a channel using a layer-1 transaction from an earlier state of the channel.
This implies that each closing transaction, $e\in E(G)$, must come with a delay, $D_e$, measured in block height, that is chosen during the creation of the channel $e$ by the two participating nodes.
Specifically, suppose that a node, $u$, broadcasts its intention to unilaterally close a channel, $e=\{u,v\}$, by confirming on the blockchain a transaction, $\tau$, that corresponds to an old state of $e$.
Then, the node $v$ can get all the funds in the channel by confirming a transaction, $\tau'$, that effectively proves that $\tau$ has expired, within a time window of $D_e$ blocks since the confirmation of $\tau$.

\subsection{The attacks}
The above inherent timing limitations of the LN protocol render it susceptible to attacks that try to force a mass-exit event that violates the delay bounds of many edges, thus forcing a deviation from the intended protocol behavior.

We study the following two attacks of this type, that are performed by an adversary that controls a small coalition of $k$ nodes:

\textbf{(1) Zombie attack:}
The adversary controls a set of $k$ nodes, $Z\subseteq V(G)$, that hold exactly one side of many channels; that is, $E' = E(G) \cap (Z\times (V(G)\setminus Z))$.
The adversary renders all channels in $E'$ unresponsive, simply by not participating in the protocol.
This can be implemented either when $Z$ is a coalition of adversarial nodes, or when an eclipse-type attack (see, e.g.~\cite{heilman2015eclipse, marcus2018low}) is performed on all nodes in $Z$.
This forces many honest nodes to submit layer-1 channel closing transactions.
We show experimentally that, under realistic layer-1 congestion conditions,
the zombie attack can cause a significant amount of funds to be locked for a large window of time.
This deviates from the protocol which attempts to impose a much smaller delay, $D$, for closing a channel.
This attack is similar to the \textit{griefing attack}: the user is forced to broadcast a transaction on Bitcoin layer-1 to unilaterally close the channel, and potentially pay a high fee due to the congestion generated by the attack.
This is an attack that does not aim at stealing funds, but rather to ``vandalize'' both layer-2 (unusable channels) and layer-1 (congestion).
The attack also causes an implicit monetary cost for routing nodes due to the loss of income from serving LN payments during the period the funds are locked.

\textbf{(2) Mass double-spend attack:}
The adversary again controls $Z$ as in the zombie attack (in contrast to the zombie attack, for a mass double-spend attack it is not enough to simply eclipse $Z$).
The adversary attempts to perform multiple double-spends by submitting closing transactions for all channels in $E'$ that correspond to earlier states of the protocol.
The honest nodes that monitor the blockchain respond by submitting penalty transactions on the blockchain, as soon as any offending transaction is confirmed.
Honest users' funds can be spent by the adversary when an honest penalty transaction fails to get confirmed on the blockchain before the delay $D$ has passed. The adversary succeeds and makes profits when funds are moved out of the channel's multi-signature address.
We experimentally show that, under realistic congestion conditions, if the watchtower algorithms monitoring the blockchain are not configured correctly, the adversary can succeed in stealing significant amounts of funds.

\subsection{Our main results}

Informally, our main findings can be summarized as follows.

\begin{itemize}
\item
\textbf{Susceptibility to vandalism by a small coalition:}
We demonstrate experimentally that the zombie attack can be used to lock a significant amount of funds for long periods of time.
Our simulations are performed 
under historically plausible congestion conditions.
The attack succeeds in forcing the funds of honest protocol participants to be locked for a duration that is much longer than the intended upper bound.
A detailed exposition of our findings 
is given in Section \ref{sec:zombie}.

\item
\textbf{Incentive-incompatibility in the presence of a small coalition:}
We demonstrate experimentally that, 
assuming an adversarial coalition of $k=30$ nodes, 
and under a plausible model for the expected profit of the adversary,
the LN is not incentive-compatible.
Our simulations are performed on a recent LN graph topology,
under historically-realistic congestion conditions,
and assuming a realistic watchtower algorithm.
Our mathematical modeling makes mild statistical assumptions on channel balances.
The full details of our mathematical modeling are given in Section \ref{sec:coalition},
and our experimental findings are presented in Section \ref{sec:mass-double-spend}.
We demonstrate experimentally that 
with the current LN topology, under historically realistic congestion conditions, using a realistic watchtower algorithm, under mild statistical assumptions on channel balances, and assuming a coalition of at most $k=30$ nodes, the LN protocol is not incentive-compatible.
In fact, under reasonable model parameters, the adversarial coalition has a strategy for stealing funds from the honest protocol participants, with an expected profit of more than 750 BTC. In this case, there is a simplification since we are considering that the business of the attackers has a volume that is less than the expected profit of the attack. Nonetheless, we believe that the \textit{trustless} nature of the network is questioned by the possibility of performing these attacks.
We emphasize that our results do not imply that \textit{any} $k$-coalition can perform this strategy, but rather that such a $k$-coalition \textit{exists} and can be computed efficiently in practice.
\end{itemize}

\subsection{Theoretical justification}
Our experimental findings outlined above suggest that the LN is potentially susceptible to mass exit attacks by a small coalition of adversarial nodes.
This phenomenon can be explained by the topological properties of the graph $G$ of the $LN$.
In particular, it has been recently observed that 
$G$ is scale-free, and follows a power-law degree distribution \cite{martinazzi2020evolving}.
This is to be expected due to the complex social dynamics that lead to the formation of $G$.
It has been shown in \cite{gast2016approximation} that in power-law degree graphs, by splitting the nodes into sets of high and low degree nodes, $V(G)=V_{\text{high}} \cup V_{\text{low}}$, we obtain a nearly-optimal max-cut.
This is true even for very small sets $V_{\text{high}}$ (depending on the power-law exponent; see \cite{martinazzi2020evolving} for precise bounds).
As we explain in Section \ref{sec:coalition},
for our attacks this implies that, if the adversary controls $Z=V_{\text{high}}$, then both attacks can be performed effectively by targeting all the channels between $V_{\text{high}}$ and $V_{\text{low}}$.

The above observation implies that the effects of the attacks that we study should be expected to get worse for large sizes of the LN, in the following sense: the worst-case $k$-coalition, for the \textit{same size}, $k$, can cause more damage in a larger network (either by causing a higher delay in closing channels or more stolen funds, depending on the attack).

\subsection{Organization}
The rest of the paper is organized as follows.
In Section \ref{sec:coalition} we show how both the zombie and the mass double-spend attacks can be modeled mathematically so that the problem of selecting a $k$-coalition of adversarial nodes that maximizes the efficiency of the attack can be expressed as a variant of the Max-Cut problem.
We also present a simple greedy heuristic for solving this problem.
In Section \ref{sec:experimental} we discuss the main statistical assumptions for our simulations.
Section \ref{sec:zombie} presents our experimental results on the zombie attack.
Section \ref{sec:mass-double-spend} presents our experimental results on the mass double-spend attack.
In Section \ref{sec:mitigations} we discuss possible measures for mitigating the attacks.
Section \ref{sec:related} reviews related work.
We conclude in Section \ref{sec:conclusions}.

\section{Selecting an adversarial coalition as a graph cut problem}
\label{sec:coalition}

We now describe our methodology for choosing an adversarial $k$-coalition, i.e. a coalition of $k$ nodes.
To that end, we first formulate an auxiliary graph-cut combinatorial optimization problem.
We then explain how this problem can be used to compute near-optimal coalitions.

\subsection{The graph cut problem}
In a graph, $G$, a \textit{$k$-lopsided cut} is a bipartition, $(Z, V(G)\setminus Z)$, of $V(G)$, with $|Z|=k$.
The \textit{$k$-Lopsided Max-Cut} problem ($k$-LMC) is defined as follows. The input consists of a graph $G$ and some $k\in \mathbb{N}$, and the goal is to find a $k$-lopsided cut, $(Z, V(G)\setminus Z)$, maximizing the number of edges in the cut, i.e.~$|E(Z,V(G)\setminus Z)|$.
Similarly, in the \textit{$k$-Lopsided Weighted Max-Cut} problem ($k$-LWMC), the input consists of a graph, $G$, with non-negative edge capacities, and some $k\in \mathbb{N}$, and the goal is to find a cut maximizing the total capacity of the edges in the cut, i.e.~$\sum_{e\in E(Z,V(G)\setminus Z)} c(e)$, where $c(e)$ denotes the capacity of $e$.

\subsection{From graph cuts to mass exit attacks}

\subsubsection{From $k$-LMC to zombie attacks}
The problem of computing a worst-possible $k$-coalition $Z\subseteq V(G)$ of adversarial nodes to perform the zombie attack is precisely the problem of computing a $k$-LMC in $G$.
This is because any edge can be attacked precisely when exactly one of its endpoints is in the adversarial coalition (an adversary cannot attack itself, and an honest node cannot attack another honest node).
Moreover, the effectiveness of a zombie attack is maximized when the average delay to close a channel is maximized.
This occurs when the number of channels under attack, i.e. the number of edges in the cut, is maximized.

\subsubsection{From $k$-LWMC to mass double-spend attacks}
\label{sec:from-k-lwmc-to-mass-double-spend}
We now argue that, under mild assumptions, the problem of computing a worst-possible $k$-coalition of adversarial nodes to perform the mass double-spend attack, can be reduced to the problem of solving $k$-LWMC on $G$.
For any channel $e=\{u,v\}\in E(G)$ under attack, fix an orientation $(u,v)$ such that $u\in Z$, and $v\notin Z$;
let $b_t(e)\in [0, c(e)]$ denote the balance of $u$ in the channel at time $t$; in particular, $b_t(e)=\alpha$ when at time $t$ node $u$ owns $\alpha$ of the capacity of $e$ and $v$ owns $c(e)-\alpha$.

The quantity $b_t(e)$ as a function of $t$ is difficult to model and predict during the attack.
This is a challenge for our modeling because the amount of funds that the attacker can steal from a channel depends on the current balance.
We avoid the issue of estimating $b_t(e)$ by introducing a mild technical assumption.
Suppose that the adversary participates honestly in the protocol for some amount of time before the attack begins.
Let $e$ be some channel where the adversary controls exactly one of its endpoints.
Assuming that funds in the LN are allocated efficiently, it follows that in any long enough window of time, we expect to observe times $t$ and $t'$  where $b_t(e) \geq  c(e)-\eps$, and $b_{t'}(e) \leq \eps$, for any $\eps>0$.
This is because, otherwise, we could remove some of the capacity in $e$ and still be able to route the same set of LN transactions, which would violate capital efficiency.
Before the attack begins, the adversary participates in the protocol honestly.
During this period, for all of its channels, the adversary collects closing transactions that give almost all the funds to the adversary. When the attack begins at time $t$, the adversary stops participating in the LN protocol.
Thus, the balances of all channels under attack remain fixed throughout the attack.
By the above discussion, it follows that if the adversary succeeds in stealing the funds from a channel, $e$, then the adversary profits a total of at least $c(e)-b_t(e)-\eps$.
On the other hand, the adversary fails to steal the funds of channel $e$ precisely when an honest node manages to get the penalty transaction confirmed before the expiration;
when this happens, the adversary gets penalized by losing all the funds in the channel, thus the adversary loses $b_t(e)$, since this is the balance rightfully owned by the attacker at the moment of the attack.
We make the following simplifying assumptions to reason about LN channels' balances:

\begin{description}
    \item{\textbf{Assumption 1:}} 
    For any channel $e$, conditioned on the adversary successfully stealing the funds from $e$, the expected profit of the adversary is at least $c(e)/2$.
\end{description}

In particular, Assumption 1 holds if we assume that when the attack happens at time $t$, we have $\mathbf{E}[b_t(e)] = c(e)/2$, setting $\eps=0$. To support Assumption 1, we performed a brief analysis of the nodes in 30-LWMC: they are well-connected nodes, and most of them seem to be sellers of inbound liquidity, routing nodes, exchanges, and wallet providers. This empirically supports the validity of Assumption 1: if many nodes in the coalition, for example, sell liquidity, this means that they open channels with their customers, and then customers use these channels to get paid to sell goods or services. Therefore, the adversarial coalition would have a favorable commitment (i.e., the first one), since they opened the channel providing the liquidity, but it is expected that, over time, these channels will be used to acquire services, so the balance of the channels will eventually go to the victim node.

The total profit of the adversary depends on the probability of stealing the funds of any channel $e$.
To that end, it is reasonable to assume that all honest nodes use the same strategy.
Therefore, by symmetry, we arrive at the following assumption:

\begin{description}
    \item{\textbf{Assumption 2:}} 
    There exists some $p>0$ such that for all $e\in E(G)$ under attack, the adversary succeeds in stealing the funds from $e$, independently, and with probability $p$.
\end{description}

Let $P_e$ denote the profit that the adversary makes by attacking $e$, and let $P=\sum_e P_e$ be the total profit of the adversary.
By Assumptions 1 \& 2 and the linearity of expectation, we get that
$\mathbf{E}[P] = \sum_{e\in E(Z,V(G)\setminus Z)} \mathbf{E}[P_e] = \sum_{e\in E(Z,V(G)\setminus Z)} (p-1/2)\cdot c(e) = (p-1/2) \sum_{e\in E(Z,V(G)\setminus Z)} c(e)$.
In other words, the expected profit of the adversary is equal to the total capacity of the computed cut, times a scaling factor of $p-1/2$.
The following proposition summarizes the above discussion.

\begin{description}
\item{\textbf{Proposition 1:}}  Under Assumptions 1--2, for any $p>1/2$, the mass double-spend attack is profitable in expectation.
Moreover, the $k$-coalition that maximizes the profit of the adversary is precisely an optimal solution to $k$-LWMC.
\end{description}

\subsection{Computing lopsided max-cuts}
Since the $k$-LMC problem is a generalization of Max-Cut, it follows that it is also NP-hard \cite{karp1972reducibility}.
Perhaps surprisingly, the generalization of Max-Cut that we consider does not appear to have been studied in this precise form previously.

We observe empirically that a simple greedy algorithm computes very good solutions.
The greedy algorithm is as follows.
We start with the cut $(\emptyset, V(G))$, which contains all vertices on the right side.
For $k$ steps we repeat the following. We greedily pick the vertex on the right side to move to the left side, maximizing the total capacity of edges in the cut in $k$-LWMC, and the number of edges in $k$-LMC.

We remark that obtaining a better solver for $k$-LMC and $k$-LWMC can only strengthen our results, since better solutions directly improve the effectiveness of the attacks.

\begin{figure}
    \centering
    \includegraphics[width=0.4\textwidth]{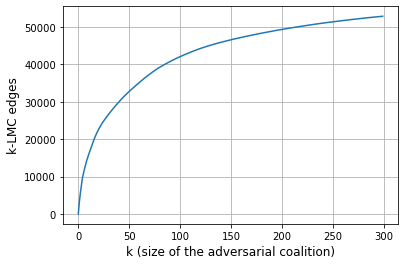}
    \includegraphics[width=0.4\textwidth]{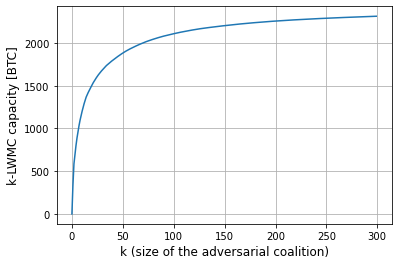}
    \caption{The value of the greedy solutions for $k$-LMC and $k$-LWMC, as a function of $k$.}
    \label{fig:lopsided}
\end{figure}

The values of the solutions for $k$-LMC and $k$-LWMC that we computed using the greedy algorithm, as a function of $k$, are presented in Figure \ref{fig:lopsided}.
The best solutions that we computed for $k$-LMC and $k$-LWMC contain respectively $\OPT = 63251$ edges and $\OPTW = 2464.37$ BTC.
Table \ref{tab:lopsided} gives the solutions for some representative values of $k$.
We observe that even for very small coalition sizes, $k$, there exist solutions that are close to the global optimum.
For example, even with $k=30$ nodes, the adversary can attack a number of channels that is $31.75\%$ of the global max-cut, which is $23.7\%$ of all channels in the network.
This phenomenon agrees with the behavior expected from the theory of scale-free networks, as discussed in the previous Section.

\begin{table}[H]
  \begin{tabular}{ccc}
    \toprule
    $k$ & $k$-LWMC capacity (\% of $\OPTW$) & \# $k$-LMC edges (\% of $\OPT$) \\
    \midrule
10 & 1199.89 (48.69\%) & 10911 (17.25\%) \\
30 & 1685.13 (68.38\%) & 20084 (31.75\%) \\
100 & 2107.70  (85.53\%) & 35447 (56.04\%) \\
300 & 2312.47 (93.85\%) & 44522 (70.39\%)\\
  \bottomrule
\end{tabular}\\
  \caption{Result of the greedy solutions computed for $k$-LMC and $k$-LWMC on the LN graph $G$.}
  \label{tab:lopsided}
\end{table}

\section{The experimental setup}
\label{sec:experimental}

We now describe the modeling of the LN used in our experiments. The graph $G$ of the LN used in our experiments was obtained using the \textit{lnd} lightning node implementation.
The snapshot of the network that we use was taken in May 2022
and contains $n=|V(G)|=17813$ nodes and $|E(G)|=84927$ channels.

\subsection{Modeling layer-1 congestion}

The effectiveness of the attacks we consider depends on the confirmation times of the transactions performed by the adversary and the honest nodes.
Therefore, the results depend on the congestion of the Bitcoin blockchain during the attack.

One way to estimate confirmation times is to use statistical models, such as \cite{gundlach2021predicting}.
However, this approach requires estimating the parameters of the models involved, which can introduce bias in the results.
For this reason, we decide to perform simulations using historical data on the Bitcoin mempool \footnote{The code implementing the simulations and the experiments can be found at \url{https://github.com/CosimoSguanci/Mass-Exit-Attacks-on-LN}}.

All of our simulations are performed under two different congestion scenarios:

\begin{description}
\item{\textbf{Scenario 1:}} The attack starts on Dec. 7, 2017 at 08:15 (CDT), which marks the beginning of a period of high congestion.
    
\item{\textbf{Scenario 2:}} The attack 
starts on Jan. 1, 2022 at 00:00 (CDT), which is during a period of typical congestion.
\end{description}

\begin{figure}[H]
\begin{center}
    \includegraphics[width=.45\textwidth]{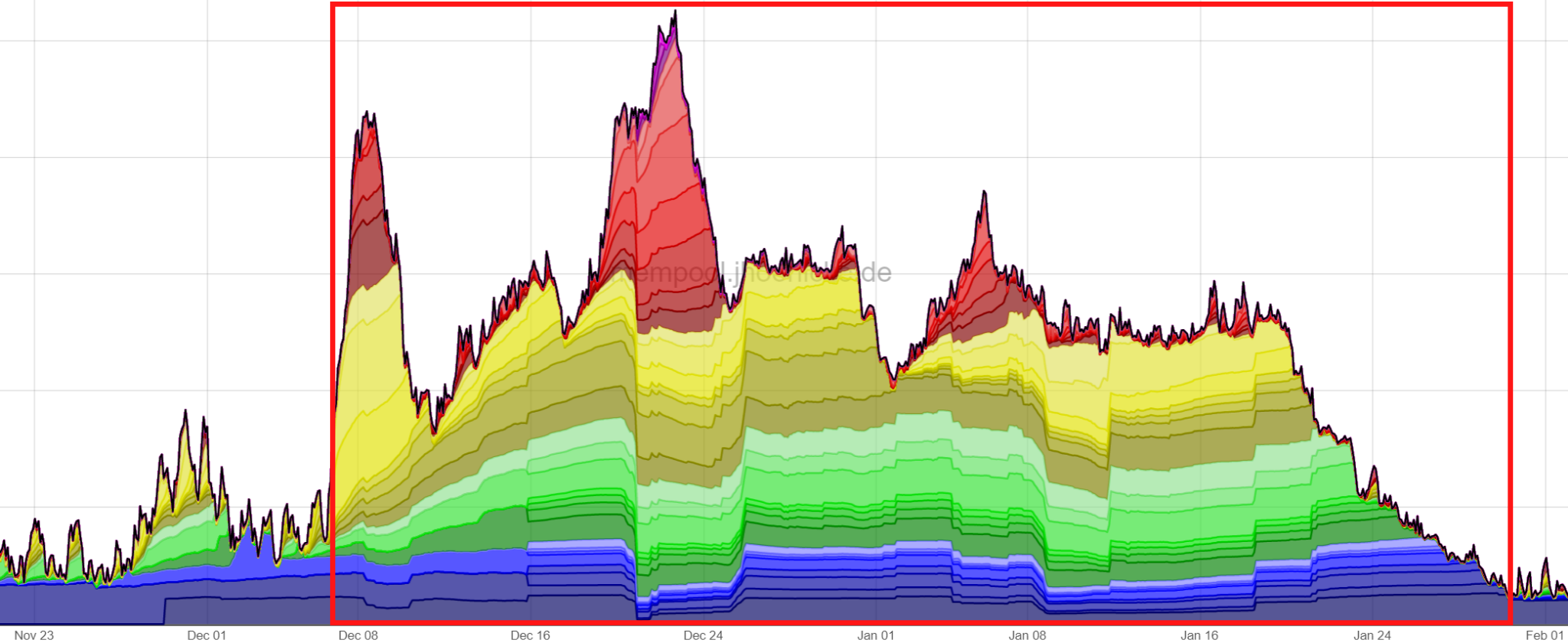}
\end{center}
    \caption{Levels of congestion between December 2017 and February 2018, showing the end of the period of congestion that started around Dec. 6, 2017. }
    \label{fig:congestion}
\end{figure}

Scenario 1 represents a worst-case situation for the victims, since the attack takes place during a period of high congestion on layer-1.
The data used to run the experiments was obtained from \cite{mempool-data}, with a snapshot taken every minute regarding the number of transactions in the mempool for several predefined fee ranges.
Figure \ref{fig:congestion} depicts the congestion during the selected period.
The mempool congestion in the selected period lasted approximately until the end of January (about 8000 blocks).

\subsection{Modeling the strategy of honest nodes}

In both the attacks presented in this paper, we consider two different strategies for honest players:

\begin{itemize}
    \item \textit{Static strategy}: 
    All honest nodes use a fixed fee when transmitting their transactions.
    This is the optimistic case for the adversary, since it is harder for honest players to get their transactions confirmed during periods of high congestion.

    \item \textit{Dynamic strategy}: 
    Honest nodes increase the fee of penalty transactions until they get confirmed or the delay expires.
    The rate of increase is controlled by three parameters: The initial fee, the \textit{step} parameter, $s$, and a parameter $\beta>1$.
    Every $s$ blocks, any honest node increases the fee of any pending transaction via replace-by-fee (RBF) or other similar mechanisms, by multiplying its fee by $\beta$.
    Therefore, the fees increase exponentially.
\end{itemize}

The dynamic strategy requires some considerations regarding the various types of on-chain transactions that both the attacker and the victims have to broadcast. First of all, penalty transactions only require the signature of honest nodes, therefore RBF can be used, and the dynamic honest strategy is feasible in the mass double-spend attack. In the zombie attack, the fee increase is related to channel force-closing transactions: channel commitments are signed by both parties when the commitment is created, therefore in general the fee is static and decided when the transaction is signed. However, with the implementation of \textit{anchor output-based channels} \cite{anchor-outputs}, the fee of force-closing transactions can be bumped, therefore the dynamic strategy of honest nodes can also be implemented during the zombie attack.
In the zombie attack, if the strategy is static, the fee will remain the same for the entire duration of the simulation, until all the zombie channels have been closed. When the dynamic strategy is adopted, the fee will be increased every $s$ blocks, for all the transactions that are still unconfirmed. For example, if $s=2$, then every 2 blocks the fee related to unconfirmed zombie channel closing transactions will be bumped.
In the mass double-spend attack, honest nodes' transactions are submitted as soon as the corresponding malicious transaction is confirmed (effectively simulating the behavior of \textit{watchtowers}). The initial fee for victims' penalty transactions is set as the average fee at the moment of submission. After submission, fee increments in the dynamic scenario follow the same mechanism as in the zombie attack simulation.

\section{Results of the Zombie Attack}
\label{sec:zombie}

\subsection{Outline of the experiments}
The simulation of the zombie attack works as follows.

\begin{enumerate}
    \item At the beginning of the simulation, there are $n$ zombie channels that must be closed. These would typically be the channels in the solution to $k$-LMC.
    \item We compute the number of transactions contained in any new block from historical blockchain data.
    For example, in scenario 1, since the time window starts on Dec. 7 2017 at 08:15 (CDT), the first block that will be mined is block \#498084.
    \item We check if there is enough space in the block for any monitored transaction by counting the number of transactions in the mempool with a higher fee (also considering transactions that were already in the mempool with the same fee rate at the moment of submission).
    \item If the block can fit some of the closing channel transactions, we remove them from the remaining number of zombie channels to be closed.
    \item When we reach a state in which all the LN channel closing transactions have been confirmed, the simulation ends.
\end{enumerate}

In the experiments related to the zombie attack, we do not consider the delay set at channel opening (\verb|to_self_delay|). When simulating scenario 2, we don't consider the historical number of transactions contained in the current block, since when the congestion is low blocks may not be full. This is not realistic since we are adding many transactions to the mempool, thus we consider that each block contains the average number of transactions per block as computed in scenario 1.

\subsection{Results}

\subsubsection{Static honest strategy}

\begin{figure*}
\begin{center}
        \includegraphics[width=.45\textwidth]{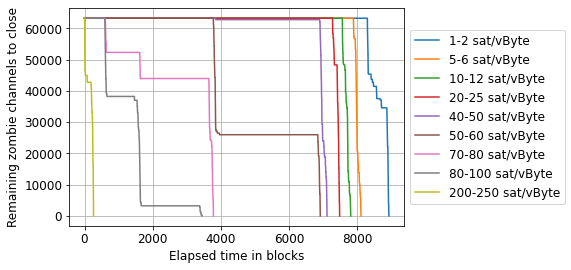}
        \includegraphics[width=.45\textwidth]{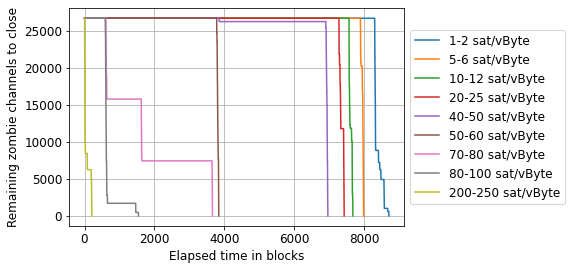}
        \includegraphics[width=.45\textwidth]{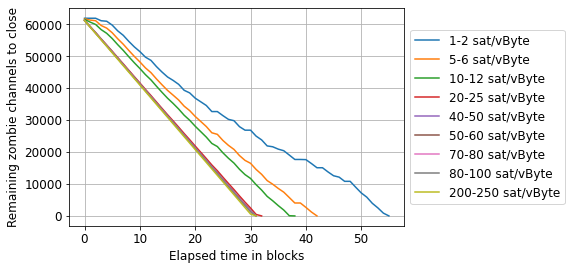}
        \caption{Remaining zombie channels to close as a function of time measured in blocks, for various fee ranges, with static honest player strategy, in scenario 1 (attacking the best LMC on the left and 30-LMC on the right) and scenario 2 (in the center).}
        \label{fig:zombie_channels_remaining}
\end{center}
\end{figure*}

\begin{figure*}
\begin{center}
    \includegraphics[width=.45\textwidth]{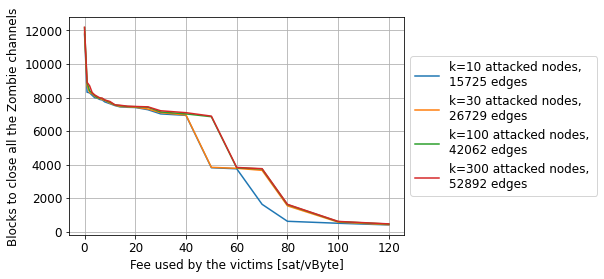}
    \includegraphics[width=.45\textwidth]{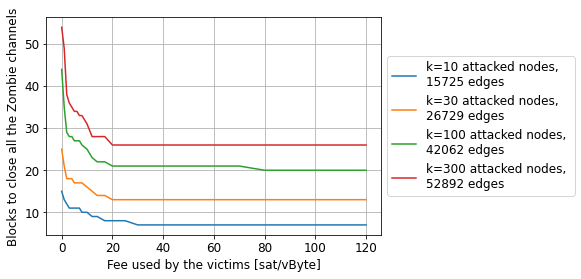}
\end{center}
\caption{Time measured in blocks needed to close all the zombie channels, as a function of the fee used by the victims for different values of $k$, with static honest player strategy, in scenario 1 (left) and scenario 2 (right).}    \label{fig:zombie_channels_remaining_various_k}
\end{figure*}

Figure \ref{fig:zombie_channels_remaining} shows the number of remaining zombie channels to be closed as a function of time (measured in blocks) during the simulation, for various fee rates used by honest nodes (that are fixed since we are in the static case). The results in the high congestion situation are consistent with the duration of the congestion: when the fee is less than 40-50 sat/vByte, all transactions tend to be confirmed approximately at the same time as congestion decreases significantly (around the 8000th block from the beginning, just under two months).
In the same figure, it is possible to notice that, even for a small coalition size ($k=30$), the zombie attack causes a high delay. The difference between attacking the best LMC and 30-LMC is not very pronounced: this is caused by the fact that in this case the layer-1 congestion is not dominated by LN transactions, but rather by historical transactions that were in the mempool in the analyzed time window. For scenario 2 (low congestion), in Figure \ref{fig:zombie_channels_remaining} we observe that the delay to close a channel is significantly smaller, consistently with respect to a scenario of typical congestion. Figure \ref{fig:zombie_channels_remaining_various_k} depicts the number of blocks needed to close all the zombie channels as a function of the fee rate used by the victims for the closing channel transactions, for different values of $k$.
We observe that even when attacking only a few nodes ($k=10$), the adversary causes damage to users in terms of the time their funds are locked, when the congestion is high.
Indeed, users are forced to pay a high fee of about 70 sat/vByte to redeem all the funds before two weeks ($\approx 2000$ blocks) since the moment of the attack.

\subsubsection{Dynamic honest strategy}

\begin{figure*}
    \begin{center}
        \includegraphics[width=.45\textwidth]{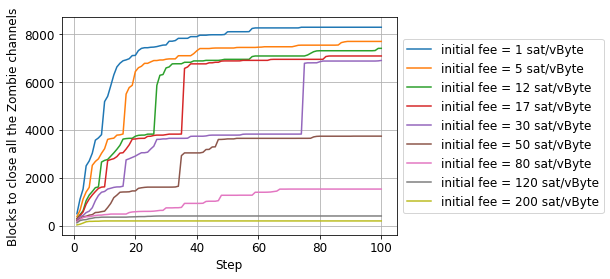}
        ~
        \includegraphics[width=.45\textwidth]{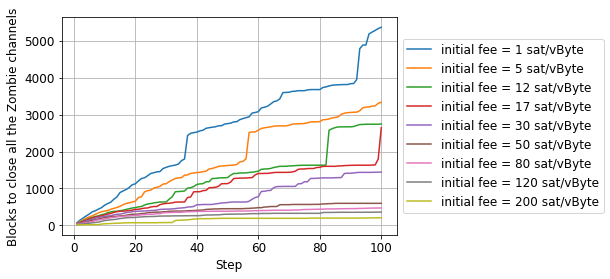}
        ~
        \includegraphics[width=.45\textwidth]{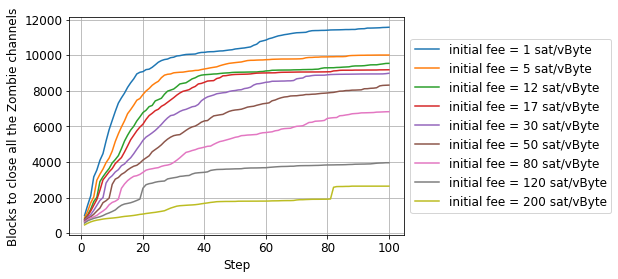}
    \end{center}

    \caption{Number of blocks needed to close all the zombie channels, with dynamic honest player strategy, as a function of the step parameter, in scenario 1, attacking 30-LMC (beta=1.01 on the left, beta=1.1 on the right) and 1 million zombie channels, in the center (beta=1.01).}
    \label{fig:zombie_channels_dynamic}
\end{figure*}

In the dynamic honest strategy case, the fee rates used by victims of the attack can increase over time in response to a delay in the confirmation of the transactions. 
We consider various values for the $\beta$ and \textit{step} parameters. Figure \ref{fig:zombie_channels_dynamic} shows the number of blocks needed to close all the zombie channels, as a function of the parameters of the dynamic strategy.
We observe that, 
even for a coalition of size $k=30$,
if honest users wish to close their channels in less than 2000 blocks, then either the initial fee must be set high (about 80 sat/vByte), or the fee must be increased aggressively. A more aggressive honest strategy, with $\beta=1.1$ leads to lower waiting times as expected, even if it causes high transaction costs for users.

Figure \ref{fig:zombie_channels_dynamic} also considers a hypothetical case with 1 million zombie channels, which corresponds to a much larger graph in a possible future snapshot of the LN.

\section{Results of the Mass double-spend attack}
\label{sec:mass-double-spend}

We now present our results on the mass double-spend attack.

\subsection{Outline of the experiments}

The simulation is performed as follows.

\begin{enumerate}
    \item When a new block is mined, we check if some of the LN transactions can be included in the block, counting the number of transactions with a higher fee and checking if this number is less than the number of transactions historically contained in the current block.
    \item If there is some potential space for LN transactions in the new block, we start by iterating over transactions: before including them into the block, we check their position in their specific fee range in the mempool. If, at the moment of submission to the mempool, there were already some transactions at the same fee rate, these transactions would need to be confirmed before those that we are monitoring.
    \item At the beginning, there are only adversarial transactions: we fit as many as we can in the first block that has available space for them. For each confirmed adversarial transaction, we submit to the mempool the corresponding honest penalty transaction. The fee rate of each penalty transaction is computed as the average fee from the mempool data at the moment of submission.
    \item For each penalty transaction, if it has not been confirmed yet and \verb|to_self_delay| blocks have been mined, the adversary submits another transaction (also called \textit{sweep transaction}) that spends the funds from the channel's multi-signature address.
    \item When all the transactions we submitted are confirmed, the simulation ends. We refer to channels for which the attacker was able to spend the stolen funds as \textit{compromised channels}.
\end{enumerate}

In this case, unlike in the zombie attack simulation, the transactions we add are submitted at different times and potentially with different fees.
Therefore, we must keep track of the position of each penalty transaction in the mempool in the corresponding fee range, to detect its confirmation block as accurately as possible.

When we decide that, for example, $n$ attacker transactions are included in a block, we are actually replacing $n$ transactions which, if we look at historical mempool data, were removed from the mempool, but in the simulation that we are executing they should still be in the mempool as unconfirmed. For simplicity, in our simulation we consider only the transactions that are directly replaced by LN transactions, and we don't consider the ``cascade'' of replaced transactions.

\subsection{Results}

\subsubsection{Static honest strategy}

\begin{figure*}
    \centering
    \includegraphics[width=.45\textwidth]{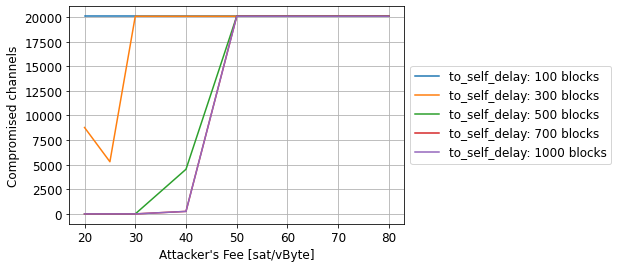}
    \includegraphics[width=.45\textwidth]{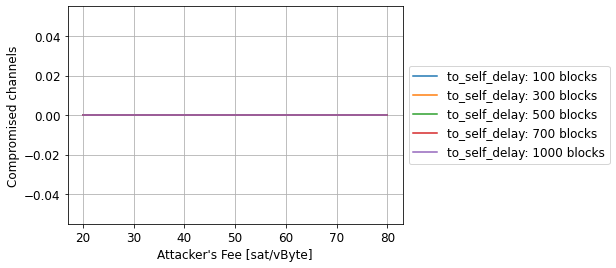}
    \caption{Number of compromised channels as a function of the fee rate used by the attacker for various values of \texttt{to\_self\_delay} with static honest player strategy, in scenario 1 (left) and scenario 2 (right), attacking channels in 30-LWMC.}
    \label{fig:mass-double-spend-static}
\end{figure*}

Figure \ref{fig:mass-double-spend-static} shows the number of compromised channels, in scenario 1, as a function of the fee paid by the attacker. In addition, the adversary uses a fixed fee rate of 100 sat/vByte for the second adversarial transaction, which tries to spend the funds from the on-chain multi-signature address.
We observe that, even for large delays (e.g.~1000 blocks), the number of compromised channels is high, especially if the attacker is willing to pay a high fee for its transactions.
We also note that it is not guaranteed that increasing the fee for the attacker leads to a better attack outcome, as it is possible to see with a delay of 500 blocks.
This is because a higher fee also implies that the penalty transaction is submitted earlier, and the effect on the results depends on the level of congestion at the moment that each victim transaction is submitted.
The level of congestion in the considered time window is not constant, it changes significantly and not monotonically over time: therefore, it could happen that using a higher fee causes penalty transactions to be submitted in a moment in which the mempool is less overloaded, hence they can also be confirmed earlier, actually reducing the effect of the attack.
Figure \ref{fig:mass-double-spend-static} also shows consistent results with respect to a typical congestion period: a small \verb|to_self_delay| is generally enough to minimize the risks for LN users.

\subsubsection{References to LN nodes implementation}
\label{sec:mass-double-spend-implementations}
For the dynamic honest strategy case, we want to fix a reasonable value for the channel dispute delay to be used during our experiments. We investigate how the lnd implementation currently handles the default delay agreed upon at channel opening, and how this affects our attack.
In the Lightning Network specification (BOLT), \verb|to_self_delay| is the parameter that sets the delay (in blocks) that the other end of the channel must wait before redeeming the funds in case of uncooperative channel closure. The default delay is scaled according to the channel capacity. Considering an average channel capacity of $0.045$ BTC (as of May 2022 \footnote{Data from \url{https://bitcoinvisuals.com/ln-capacity-per-channel}}), the average default delay would be $\approx 539$ blocks. For the dynamic honest strategy experiments, we will consider this value for the delay.

\subsubsection{Dynamic honest strategy}

\begin{figure}
    \centering
    \includegraphics[width=.45\textwidth]{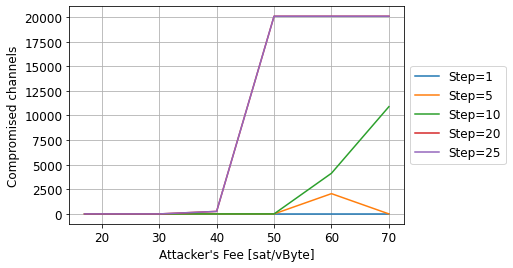}
    \caption{Number of compromised channels as a function of attackers' fee rate
    with dynamic honest player strategy, for various values of the step parameter (scenario 1), attacking channels in 30-LWMC.}
    \label{fig:mass-double-spend-dynamic-scenario1}
\end{figure}

The dynamic case, in which victims are also increasing their fee if the penalty transaction is not confirmed after the number of blocks determined by the \textit{step} parameter, is shown in Figure \ref{fig:mass-double-spend-dynamic-scenario1}. Similar to the zombie attack, the study of the dynamic case is useful to propose possible strategies to defend against this attack. 
As a remark, the curves shown in Figure \ref{fig:mass-double-spend-dynamic-scenario1} are not monotonic, because the congestion level in the considered time window is not constant. The results show that an aggressive honest strategy (\textit{step} $\leq5$) is needed to avoid significant risks for users' funds. Since the initial honest fee rate is computed as the average fee at the moment of submission to the mempool, and each victim's transaction is sent when the corresponding attacker's transaction is included in a block, even using a higher fee doesn't necessarily always optimize the attacker's strategy.

\subsubsection{Considerations regarding attackers' profit}
\begin{figure*}
\begin{center}
        \includegraphics[width=.5\textwidth]{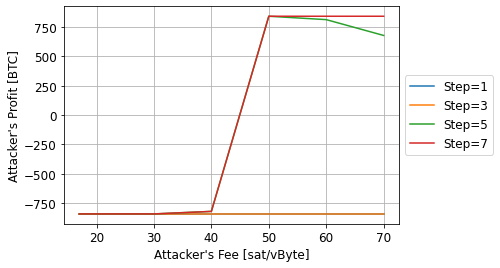}
        \includegraphics[width=.4\textwidth]{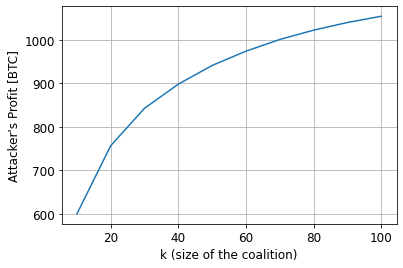}
        \caption{Attacker's profits as a function of the initial fee used by the attacker in scenario 1, attacking 30-LWMC, with dynamic honest and adversarial strategy, for various values of the step parameter (on the left). On the right, the attacker's profit as a function of $k$ is shown.}
        \label{fig:attacker_profits}
\end{center}
\end{figure*}

In a realistic setting, some of the penalty transactions submitted by the victims may be confirmed before the expiration of the \verb|to_self_delay|. These transactions will cause a penalty for the attacker, as the balance of the channel will be claimed by the victim.
Following our previous assumptions, the total expected profit can be computed as follows:
$$\mathit{profits} = c/2 \cdot n - (c/2 \cdot (a - n)) $$
\begin{itemize}
    \item $c$ is the average capacity per channel
    \item $n$ is the number of compromised channels
    \item $a$ is the number of attacked channels
\end{itemize}
Therefore, the whole attack is profitable when the attacker steals funds from a set of channels that contain at least half of the total capacity of the channels under attack.
The profit of the attacker under different conditions is shown in Figure \ref{fig:attacker_profits}. In the first experiment, also the adversary adopts a dynamic fee strategy for sweep transactions, starting from a fee rate of 100 sat/vByte.
We observe that even with a coalition of $k=30$ nodes, under heavy congestion, 
using an adversarial transaction fee rate of 50 sat/vByte,
when the honest nodes bump their fee every 7 blocks (about every 70 minutes), the adversary has a total expected profit of more than 750 BTC. The second experiment illustrated in Figure \ref{fig:attacker_profits} shows that, with \textit{step} = 5, an attacker's fee rate range of 50-60 sat/vByte, with dynamic honest and adversarial strategy, a coalition of 80 nodes has an expected profit of more than 1000 BTC.

\section{Mitigations}
\label{sec:mitigations}

\subsection{Mitigations for the mass double-spend attack}

\subsubsection{Increase \texttt{to\_self\_delay}}
As we observed in the simulation results, increasing the \verb|to_self_delay| helps in reducing the damage made by the attack. However, this comes at a high cost in terms of usability: if the other party is unresponsive, an honest user has to wait for that increased delay to be able to recover and use the funds locked in the channel. We note also that this modification would worsen the delay in the zombie attack.

\subsubsection{Watchtowers}

Watchtowers are services that offer protection against fraudulent commitments to prevent the loss of funds when users are offline. They act in response to a non-valid commitment posted on-chain, by broadcasting a penalty transaction as soon as the adversarial transaction has been confirmed. The exact strategy of the watchtower should be carefully decided by the operator:
our simulations suggest that during periods of high layer-1 congestion, an aggressive dynamic fee strategy should be adopted.

\subsubsection{Avoid waiting until the attacker transaction is confirmed}
A different watchtower strategy would be to monitor the mempool for adversarial transactions.
In this case, when an adversarial transaction is detected in the mempool, the watchtower attempts to \textit{front run} the adversary by submitting a channel force-closing transaction with a higher fee.

\subsubsection{On-chain congestion detection}
Authors of \cite{lotem2022sliding} discuss a mechanism to detect congested periods in the blockchain. They achieve this on Ethereum with a smart contract that makes use of the \textit{base fee} and a new opcode. A challenge-response protocol adjustment is proposed to extend the time users have to react to fraudulent behavior in case high congestion is detected.

\subsection{Mitigation for both attacks}

\subsubsection{Incentivize low-degree nodes}
Our attacks can be effective even in the presence of small adversarial coalitions.
This is because the LN network has a power-law degree distribution, which implies that there exist large $k$-LMCs and $k$-LWMCs, for small values of $k$.
If LN routing algorithms were to give priority to paths that use low-degree nodes, then it is reasonable to expect that the degree distribution of the LN would become closer to uniform, and thus the sizes of the maximum $k$-LMCs and $k$-LWMCs should decrease.
This would render both attacks less effective, for the same coalition size.
However, it is unclear how this modification would affect the routing properties of the LN.

\section{Related Work}
\label{sec:related}

LN attacks that exploit layer-1 congestion have already been described in the literature: \cite{harris2020flood} presents the \textit{Flood \& Loot} attack. This attack makes use of multi-hop payments: the attacker controls two nodes that are not directly connected, the source and the target. The paths between them are flooded with a high number of low-value HTLC payments from the source to the target. When all the HTLC payments have been completed, the source node refuses to resolve the payments with the first node in the payment path, which is the victim. The victim is therefore forced to unilaterally close the channel since the source node is unresponsive. The high number of unresolved HTLCs and the consequent congestion on layer-1 cause some of the victim transactions not to be confirmed in time, allowing the attacker to steal some of the funds.
The exploited mechanisms partially differ from those used by the attacks presented in this paper, since we do not make use of multi-hop payments, but rather direct channels between nodes. 
Similar results are obtained in \cite{mizrahi2021congestion}: however, in this work they target the entire network, and an attack that is similar to the zombie attack is described. A greedy algorithm is designed to select the channels to be flooded to maximize the locked capacity. This is also similar to what is proposed in \cite{rohrer2019discharged}, where they analyze attack strategies by making use of the topological properties of the LN: for example, they present some strategies to choose nodes to be removed for partitioning attacks.
A well-known attack on the LN is the \textit{griefing attack} \cite{mazumdar2020griefing}. In this case, the attacker saturates HTLC multi-hop payments path and makes the channels that constitute the path unusable for new payments. The corresponding channels must be force-closed on-chain, therefore users' funds are locked for the delay specified at channel creation. In general, griefing attacks do not aim to steal funds, but users are forced to pay high fees to close the channels.
The \textit{time-dilation attacks} described in \cite{riard2020time} aim at eclipsing LN nodes to delay the time at which they become aware of new blocks. This attack targets mechanisms that rely on timely reactions, such as the LN penalty mechanism. To make detection difficult the attacker exploits the random arrival of new Bitcoin blocks, that is a Poisson process.
Finally, mass exit attacks on Plasma are studied in \cite{dziembowski2021lower}.

\section{Conclusions}
\label{sec:conclusions}

In this paper, we have analyzed two attacks that exploit congestion on the Bitcoin blockchain to cause damage to the Lightning Network. Using historical mempool data related to periods of high and low transaction congestion, we have simulated two attacks: in the case of the zombie attack, funds remain locked for many blocks before users can retrieve them, while, in the mass double-spend attack, funds are at risk of being stolen.

We presented theoretical justifications for the effectiveness of the attacks, even when the adversary controls only a very small coalition.
This is based on the observation that the LN network is scale-free, and scale-free graphs have large $k$-lopsided cuts, for small values of $k$.
This suggests that our attacks should become more effective for larger network sizes, for the same coalition size, $k$.
In particular, the average expected profit of a node participating in the worst-possible $k$-coalition should grow in the future with network size.

Our simulation results suggest that watchtower algorithms should be configured carefully.
Ideally, they should monitor layer-1 congestion, and respond aggressively in the case of high congestion. We remark that our findings warn against the risks of LN centralization, rather than posing an immediate threat to users.

We leave as future work the study of both attacks under a more detailed and accurate model that incorporates transaction fees in the costs of the participants.

\bibliographystyle{IEEEtran}
\bibliography{main}

\end{document}